\documentstyle[12pt]{article}
\date{}
\begin{document}
\title{{\LARGE\sf Equilibrium Pure States and Nonequilibrium Chaos}}
\author{
{\bf C. M. Newman}\thanks{Partially supported by the 
National Science Foundation under grant DMS-95-00868
and DMS-98-02310.}\\
{\small \tt newman\,@\,cims.nyu.edu}\\
{\small \sl Courant Institute of Mathematical Sciences}\\
{\small \sl New York University}\\
{\small \sl New York, NY 10012, USA}
\and
{\bf D. L. Stein}\thanks{Partially supported by the 
U.S.~Department of Energy under grant DE-FG03-93ER25155
and by the National Science Foundation under grant DMS-98-02153.}\\
{\small \tt dls\,@\,physics.arizona.edu}\\
{\small \sl Depts.\ of Physics and Mathematics}\\
{\small \sl University of Arizona}\\
{\small \sl Tucson, AZ 85721, USA}
}
\maketitle
\begin{abstract}
We consider nonequilibrium systems such as the Edwards-Anderson
Ising spin glass at a temperature where, in equilibrium,
there are presumed to be (two or many) broken symmetry pure
states.  Following a deep quench, 
we argue that as time $t \to \infty$, although the system is usually
in some pure state locally, 
either it {\it never\/} settles permanently 
on a fixed lengthscale 
into a single pure state, or 
it does but then the pure state depends on {\it both\/} 
the initial spin configuration {\it and\/} 
the realization of the stochastic dynamics.
But this latter case can occur only if there exists an uncountable number 
of pure states (for each coupling realization) 
with almost every pair having zero overlap.
In both cases, almost no initial spin configuration is in the
basin of attraction of a single pure state; that is, the configuration
space (resulting from a deep quench) is all boundary (except for
a set of measure zero).  We prove that the former
case holds for deeply quenched $2D$ ferromagnets.
Our results raise the possibility that even if more
than one pure state exists for an infinite system, time averages don't
necessarily disagree with Boltzmann averages. 
\end{abstract}
\small
\normalsize
\section{Introduction}
\label{sec:intro}

In this paper we study nonequilibrium dynamics of spin systems.  
While our approach is general, and covers
both ordered and disordered, Ising and non-Ising systems, in
the presence or absence of a magnetic field, we will
for specificity here focus mostly on 
the dynamics of the Edwards-Anderson$^{\cite{EA}}$
Ising spin glass in zero field.  
We make no assumptions about the real- or state-space structure
of the low-temperature spin glass phase, 
but instead derive
several general principles and then explore
their consequences. 

Our results indicate that equilibrium pure state structure plays  
an important role in nonequilibrium dynamics.
E.g., we will show that (at fixed temperature) a system with many pure states 
may have very different dynamical behavior than one with only 
a single pair\footnote{We emphasize that 
we are talking here about the equilibrium pure states
present {\it at\/} the temperature at which the dynamics
is observed, as distinguished from the proposal 
(see, e.g., Ref.~\cite{HLOOV})
that metastable states with $O(1)$ free energy barriers
affecting dynamics at a given temperature are the precursors of 
equilibrium pure states at lower temperatures} --- so nonequilibrium dynamics
can serve as an important probe of the equilibrium pure state structure.

That the number of and relationships among pure states can affect
nonequilibrium dynamics may seem surprising in light
of a general supposition (see, for example, Refs.~\cite{BY,MPV}
and many of the references therein) that the time evolution of
an infinite system is confined within a single pure state for all finite
times.  A consequence of this 
supposition 
is the prevalent notion that, if there exists
at some temperature $T$ more than one pure state (e.g., due to broken symmetry),
then necessarily the 
limits $t\to\infty$ (time) and $N\to\infty$ (system size) do not
commute (when, say, measuring the state of some observable).
An equivalent statement is that time averages (as performed in
the lab) and Boltzmann averages (as performed on paper or the computer)
will give differing results.  (One proposal$^{\cite{HJY}}$ 
for avoiding this problem in theoretical
treatments of spin glass dynamics is to start with a Boltzmann
distribution of initial configurations, 
rather than a single one.)  We will argue, however, 
that these are not {\it necessary\/} consequences of broken symmetry 
or multiple pure states; they may be true in some cases, 
but not in general.  We examine here in which contexts they are true
and in which they are not\footnote{Except in some 
comments about other papers, 
we will not refer to metastable states,
often proposed as responsible for the anomalous
dynamical behavior of spin glasses.  While we have no argument
with this, we question the usefulness of the usual practice of
inserting metastability by hand, requiring a guess as to
the structure (usually in state space) and nature of the metastable states.
In the treatment presented here, metastability
may emerge naturally.}.  We begin by constructing
dynamical probability measures on spin configurations, which
will enable us to clarify notions such as time evolution within
a pure state.

\section{Dynamical Measures}
\label{sec:measures}

We will mostly, but not exclusively, consider the EA Ising spin glass 
in zero field.  Its Hamiltonian is given by:
\begin{equation}
\label{eq:EA} 
{\cal H}_{\cal J} = -\sum_{<xy>} J_{xy}\sigma_x \sigma_y\, ,      
\end{equation}
where the sites $x,y\in {\bf Z}^d$ and the 
sum is taken over nearest neighbors only.  The 
couplings $J_{xy}$ are independent random variables,
whose common probability density is 
symmetric about zero; we let ${\cal J}$
denote a particular realization of the couplings.

We consider Glauber dynamics of an {\it infinite\/} EA Ising 
spin glass starting 
from an initial (infinite-volume) spin configuration $\sigma^0$.
We regard $\sigma^0$ as chosen from the (infinite temperature)
ensemble where the individual spins are independent random variables
equally likely to be $+1$ or $-1$.
This corresponds to the experimental situation following
a deep quench, and also (for smaller systems) most numerical 
simulations.  We denote by $\omega$ a given realization
of the dynamics.  So there are three sources of randomness in the
problem, with realizations given by ${\cal J}$, 
$\sigma^0$, and $\omega$.  All three are needed to determine
the spin configuration at time $t$ (we take the starting
time to be 0). From here on we take ${\cal J}$ to be fixed.
We note that the exact choice of spin flip rates plays no role
in our analysis at positive temperature, as long as detailed
balance is satisfied.  At $T=0$ we consider only the widely used
dynamical rule where flips that are energy lowering (or
neutral or raising) occur at rate $1$ (or $1/2$ or $0$).
In all cases, $\omega$ can be regarded in the usual way
as a collection of random
times $(t_{x,i}:\,x \in {\bf Z}^d,\,i=1,2,\dots)$ when spin flips
are considered (forming a Poisson process for each $x$)
along with uniformly distributed random numbers $u_{x,i}$ that determine if
the flips are taken. 

We now define a dynamical 
probability measure $\nu_{t^*,\tau(t^*)}$ on the spin
configurations, for $0 \le \tau(t^*) \le t^*$.  
Given some ${\sigma^0}$,
we let the system evolve according to some $\omega$
up to a time $t^*-\tau(t^*)$, after which we average over
the dynamics up to time $t^*$.  That is, if we denote the dynamical
(Markov) process as $\sigma^t = \sigma^t(\omega)$ for $t \ge 0$,
then $\nu_{t^*,\tau(t^*)}$ is the conditional distribution
of $\sigma^{t^*}$ conditioned on (${\cal J}$ and ${\sigma^0}$ and)
all $(t_{x,i},u_{x,i})$'s with $t_{x,i} \le t^*-\tau(t^*)$.
So $\nu_{t^*} \equiv 
\nu_{t^*,t^*}$ represents a complete
averaging over the dynamics (and corresponds to the usual
dynamical measure---i.e., to the distribution of $\sigma^{t^*}$
for fixed ${\cal J}$ and ${\sigma^0}$), 
while $\nu_{t^*,0}$ represents no averaging
at all (and hence a single spin configuration).  To avoid awkward
notation, we generally suppress the dependence, which is understood, 
of $\nu$ on ${\cal J}$, $\sigma^0$, $\omega$ (up to time $t^* - \tau$), 
and $T$. We also note that neither $t^*$ nor $\tau$ 
depend on ${\cal J}$, $\sigma^0$ or $\omega$.
In Section~\ref{sec:fixed} we will 
briefly discuss the construction of measures 
based instead on time averaging for {\it fixed\/} dynamics.

We can now begin to 
answer the question, what does it mean for the system to
evolve or settle 
into (or ``spend all its time inside'') a single pure state?  
Consider the cube $\Lambda_L$ of linear size
$L$ and volume $|\Lambda_L|$ (which may be arbitrarily large)
centered at the origin. When $T>0$ (but not at $T=0$),
we expect that for sufficiently large $t^*$
(depending on $L$) and for almost every $\sigma^0$, the
measure $\nu_{t^*}$ approximates a (possibly mixed, possibly 
$t^*$-dependent) Gibbs state {\it restricted to the cube $\Lambda_L$\/}. 
By this we mean that
there is some (infinite volume) Gibbs state $\rho_{t^*}$
such that for any $L$ and any spin configuration $\sigma^{(L)}$
in $\Lambda_L$, the probability assigned to $\sigma^{(L)}$
by the dynamical measure $\nu_{t^*}$ and that
assigned by the equilibrium Gibbs measure $\rho_{t^*}$ 
approach each other as $t^*\to\infty$;
that is, $\nu_{t^*}(\sigma^{(L)}) - \rho_{t^*}(\sigma^{(L)})\to 0$
as $t^* \to \infty$.
More generally, we expect that for almost every
$\sigma^0$ and $\omega$, $\nu_{t^*,\tau(t^*)}$ approximates some
Gibbs state $\rho_{t^*,\tau(t^*)}$ 
providing only that $\tau(t^*) \to \infty$.

The notion that, as $t^*$ increases, $\nu_{t^*,\tau(t^*)}$ is
increasingly well approximated by some
infinite volume Gibbs state (possibly depending on $t^*$), may seem 
surprising --- especially in view of frequent assertions
(see, e.g., Ref.~\cite{CK}) that equilibrium states are 
of little relevance for the nonequilibrium dynamics of infinite systems. 
In fact, this notion is nothing more than the property that for any $L$ and any
two spin configurations $\sigma^{(L)}, \sigma'^{(L)}$, defined
within $\Lambda_L$ and agreeing on its (internal) boundary,
the ratio $\nu_{t^*,\tau}(\sigma^{(L)})/\nu_{t^*,\tau}(\sigma'^{(L)})$
converges to the usual {\it finite\/} volume Gibbs expression
coming from the EA Hamiltonian.  This kind
of convergence may be expected as it is similar to the
conjectured property of Glauber dynamics that {\it only\/} Gibbs
states are stationary. 

To answer the question posed above about the meaning of settling 
into a single pure state,
we now note that if $\tau(t^*)$ also grows sufficiently slowly,
then $\nu_{t^*,\tau(t^*)}$ should\footnote{The 
tracking of $\nu_{t^*,\tau(t^*)}$ by the {\it pure\/} state
$\rho^{\alpha(t^*)}$ is in general weaker than the tracking
by $\rho_{t^*,\tau(t^*)}$ in that 
$\alpha(t^*)$ may not have a limit and 
$\nu_{t^*,\tau(t^*)}(\sigma^{(L)}) - \rho^{\alpha(t*)}(\sigma^{(L)})$
may tend to zero only in probability rather than for almost
all $\omega$'s.  We do expect, however, that for 
almost all $\omega$'s, $\rho^{\alpha(t^*)}$
tracks $\nu_{t^*,\tau(t^*)}$ for {\it most\/} large $t^*$'s
(i.e., when domain walls are far from the origin).} (for most
$\omega$'s) approximate a {\it pure\/} (i.e., extremal) Gibbs state
$\rho^{\alpha(t^*)}$ depending on
$\sigma^{0}$ and $\omega$ (up to time $t^* -\tau$).  
If on every fixed (and arbitrarily
large) lengthscale $L$ this $\rho^{\alpha}$ eventually becomes independent
of time (after a timescale depending on $L$), then
the system has settled into the pure state $\alpha$.

We now present our main results.  Unless otherwise indicated, all 
are for $T>0$ and are independent of space dimension.  Our first
result concerns the fully averaged dynamical measure $\nu_{t^*}$.

\section{Evolution of the Dynamical Measure}
\label{sec:evolution}

{\it Theorem 1.\/} Given some ${\cal J}$, assume that for
almost every $\sigma^0$, $\nu_{t^*}$ converges
to a limiting {\it pure\/} Gibbs state $\nu_{\infty}$ as $t^*\to\infty$.  
Then $\nu_{\infty}$
is the same pure state for almost every $\sigma^0$.

\medskip

{\it Proof.\/}  We use a coupling argument where a single 
$\omega$ is used with two
starting spin realizations $\sigma^0$ and $\sigma'^{\,0}$
that differ at only finitely many sites.  Then, by the nature of
Glauber dynamics, there is a positive probability (with respect
to the $\omega$'s) that the two spin configurations merge
after a finite time.
Let $\theta>0$ represent the probability that
the difference in configurations disappears by
time $t_0$.  So for $t^* \ge t_0$,
\begin{equation}
\label{eq:one}
\nu_{t^*}^{\sigma^0}=\theta\mu_{t^*}+(1-\theta)\tilde\mu_{t^*}\, ,\,
\nu_{t^*}^{\sigma'^0}=\theta\mu_{t^*}+(1-\theta)\tilde\mu'_{t^*}\,,
\end{equation}
where $\mu_{t^*}$, $\tilde\mu_{t^*}$, and $\tilde\mu'_{t^*}$ are some
probability measures. 
It follows that for $A$ any (measurable) set of spin configurations,
\begin{equation}
\label{eq:coupling}
|\nu_{t^*}^{\sigma^0}(A) - \nu_{t^*}^{\sigma'^0}(A)|\, \le \,1-\theta\,.
\end{equation}
The same is true with ${t^*}$ replaced by $\infty$, first for
$A$ a locally defined event and then, by approximation, for general $A$.
The strict positivity of $\theta$ then 
implies that $\nu_{\infty}^{\sigma^0}$ and
$\nu_{\infty}^{\sigma'^0}$ cannot be mutually singular measures
(i.e., living on completely disjoint regions of configuration space)
and hence$^{\cite{Georgii}}$, as pure states, they must be identical.
So a change of finitely many spins in $\sigma^0$ doesn't change 
$\nu_{\infty}^{\sigma^0}$, and we can then use the 
Kolmogorov zero-one law$^{\cite{Feller}}$ to conclude
that the final pure state must be independent of $\sigma^0$. $\, \diamond$

\medskip 

Despite the innocuous look of the theorem,
it has important consequences.  Its conclusion is obvious
if there exists only one pure state, but it applies equally
to the situation where many pure states 
exist. Of course, it could logically be that only one pure
state is ``present'' (in the sense that the conclusion of the
theorem is valid) while other pure states exist but are not
physically relevant in our dynamical context. (This might even be
the case, for example, in the EA spin glass with a small nonzero field.)
But if the conclusion of the theorem is not valid,
then only one of two possibilities can occur\footnote{Logically, there 
is a third possibility that $\nu_{\infty}$ is not
a Gibbs state, but this would violate the expected behavior discussed in
Section~\ref{sec:measures} and hence we disregard it (except when
$T=0$ --- see Section~\ref{sec:lne} and Ref.~\cite{NN}).}:  either 
(1) $\nu_{\infty}$ is a mixed Gibbs 
state (which may or may not depend on $\sigma^0$),
or (2) $\nu_{t^*}$ does not converge as $t^* \to \infty$.
We note that the latter case is already 
known to occur$^{\cite{FIN}}$ in some
$1D$ disordered ferromagnets at $T=0$; on the other hand, if 
$\nu_{\infty}$ exists and does not depend on $\sigma^0$, then it
would be analogous to (and perhaps the same as)
$\rho_{\cal J}$, the average over the metastate
discussed in Ref.~\cite{NS97}.

Let's consider each of these two 
possibilities, taking into account that 
$\nu_{t^*}$ is the average over $\omega$'s 
of $\nu_{t^*,\tau(t^*)}$.  Possibility (1) implies that 
although $\nu_{t^*,\tau(t^*)}$ 
(with properly chosen $\tau$) is approximately a pure state 
$\rho^{\alpha(t^*)}$, that pure state 
depends not only on $\sigma^0$ (as expected) but also on $\omega$.
This allows (but doesn't require --- see 
Subsection~\ref{subsec:ctd}) the system always to
``land'' in a pure state in the sense that
$\nu_{t^*,\tau(t^*)} \to \rho^{\alpha(\sigma^0,\omega)}$ ---
but then the pure state is (almost) never determined solely by $\sigma^0$.

Now, the basin of attraction of a pure state $\bar{\alpha}$
may be defined as the set of $\sigma^0$'s such that
$\alpha(\sigma^0,\omega) = \bar{\alpha}$ for almost every $\omega$
(see Ref.~\cite{vEvH} for related discussions).
We claim that if $\sigma^0$ is in the basin of attraction 
of some pure state, then by a modified version (see below)
of the coupling argument 
used in the proof of Theorem 1,
the same will be true after a change of finitely many spins in 
$\sigma^0$ and so, by the Kolmogorov zero-one law, the set of
$\sigma^0$'s that are in some basin of attraction has probability
either zero or one. 
Therefore, if many pure states
are present (i.e., if $\alpha(\sigma^0,\omega)$ is not 
the same for almost all
$\sigma^0$ and $\omega$), then 
{\it the union of all their basins of attraction must form a set
of measure zero in the space of $\sigma^0$'s\/}; 
i.e., the configuration space resulting from 
a deep quench is all ``boundary'' 
in the sense that almost every initial configuration will land in one of 
several (or many) pure states depending on the realization of the
dynamics (if it lands at all).

The modified coupling argument is as follows. Let $\sigma^0$ 
and $\sigma'^{\,0}$ be fixed and let $D$ denote the finite set of
$x$'s where they differ. Rather than using the same $\omega$
for the coupled processes, we take an $\omega = (t_{x,i},u_{x,i})$
and an $\omega' = (t'_{x,i},u'_{x,i})$ that are identical for
$x$ outside $D$ but for $x$ inside $D$, they are identical 
only for times after $\sigma^t$ and $\sigma'^{\,t}$ merge. For
earlier times, $\omega$ and $\omega'$ inside $D$ are independent
of each other. If $A'$ is an event defined in terms only of 
$\omega'$ with Prob$(A') > 0$ and $M$ denotes the event of
eventual merger of the two processes, then since $\omega$ inside
$D$ may (with small but strictly positive probability) force
a merger by a small time $\varepsilon$, it follows that
Prob$(A' \cap M) > 0$. Assuming 
$\alpha(\sigma^0,\omega) = \bar{\alpha}(\sigma^0)$ for almost every
$\omega$, we take $A'$ to be the complement of the event 
that $\alpha(\sigma'^{\,0},\omega') = \bar{\alpha}(\sigma^0)$
so that Prob$(A' \cap M) > 0$ would yield the contradiction
that Prob$(\alpha(\sigma^0,\omega) = \bar{\alpha}(\sigma^0)) < 1$.
It follows that  Prob$(A') = 0$ and so 
$\alpha(\sigma'^{\,0},\omega') = \bar{\alpha}(\sigma^0)$
for almost every $\omega'$, which is exactly the claim made in the
previous paragraph.

It has been speculated$^{\cite{KL}}$ that 
slow relaxation in spin glasses may be due
to points in (high-dimensional) state 
space always being ``near'' a boundary.
What we've shown here differs in fundamental respects:  our conclusion
is that almost every point in state 
space is actually {\it on\/} a boundary,
and therefore the dynamical consequences are not restricted
to very low temperatures.

We note finally that Theorem~1 may be relevant to damage
spreading$^{\cite{Bag,Grass,JR}}$, where one asks whether the damage
(i.e., discrepancy) between $\sigma^t$ and $\sigma'^{\,t}$ (with a single 
$\omega$) grows as $t \to \infty$. Theorem~1
suggests that if damage spreading 
occurs, then $\nu_{t^*}$ doesn't converge to a single pure state   
(e.g., it might converge to a mixed state, as above).

\section{Local Non-Equilibration} 
\label{sec:lne}

Before discussing possibility (2), let us
consider the physical picture
implied by Theorem 1.  Roughly speaking, some time 
after an initial quench the system
will form domains, whose average size increases 
with time, corresponding to the
different pure states.  This scenario has been analyzed for the
two-state droplet picture$^{\cite{FH,KH,TH}}$.  It is also
a well-known scenario for coarsening in a ferromagnet following a 
deep quench$^{\cite{Bray}}$. (Of course, in contrast
to the spin glass case, one {\it does\/} know
how to prepare a ferromagnet in a pure state;
for a general discussion, see Ref.~\cite{Palmer}.)

In Ref.~\cite{Bray} it was stated that the infinite 
homogeneous ferromagnet never
reaches equilibrium in any finite time (following a deep
quench) because the domain sizes (in this case, of positive
and negative magnetization) increase with time but are never
infinite on any finite timescale.  We do not consider this 
by itself to be nonequilibration because it does not
preclude the possibility that on any {\it finite\/} lengthscale, 
the system equilibrates after some 
finite time, in the sense that after that time
domain walls cease to move across the region.  
Instead, we now propose a much 
stronger version of nonequilibration --- the possibility
of {\it local non-equilibration\/} (LNE) 
on any {\it finite\/} lengthscale, which is implied 
by possibility (2) and could also occur with
possibility (1).  We will discuss the difference between
these two cases of LNE in Subsection~\ref{subsec:ctd},
but for now will not distinguish between the two.
We will also discuss below the relation between LNE and
persistence exponents$^{\cite{BDG,DBG,DG,MBCS}}$. 

By LNE we mean that in any fixed finite region, the system
never settles down into a pure state.  That is, domain
walls do not simply move farther from the region as time progresses, but
continually return and sweep across it, changing
the state within. More precisely, LNE is said to occur
unless there is some choice of $\tau(t^*)$ such that for
almost all $\sigma^0$ and $\omega$, 
$\nu_{t^*,\tau(t^*)} \to \rho^{\alpha(\sigma^0,\omega)}$, 
a {\it pure\/} state.  If LNE occurs, it would force us to revise
the usual dynamical definition$^{\cite{BY}}$ of, e.g., the EA order
parameter.  It could also mean that, for infinite systems, 
time averages and Gibbs averages could agree, despite the presence
of many pure states.  We will return to this in Section~\ref{sec:fixed}
after investigating LNE in more detail by means of the next theorem, 
which applies to both homogeneous and disordered systems.

\medskip

{\it Theorem 2.\/}  If only a single pair 
(or countably many, including a countable infinity) of pure
states exists (with fixed ${\cal J}$) 
and these all have nonzero EA order parameter, then LNE occurs.

{\it Proof.\/}   Suppose that there exists 
at $T$ (and for the given ${\cal J}$) only a single 
pair of pure states, and
assume that LNE does {\it not\/} occur so that for each 
${\sigma^0}$ and $\omega$,
there is a limiting pure state $\alpha(\sigma^0,\omega)$.  
The overlap of $\alpha(\sigma^0,\omega)$ 
and $\alpha(\sigma'^0, \omega')$ is
\begin{equation}
\label{eq:overlap}
Q({\cal J},\sigma^0,\omega,\sigma'^0, \omega') \,=\,
\lim_{L \to \infty} |\Lambda_L|^{-1} \sum_{x \in \Lambda_L}
<\sigma_x>_{\alpha(\sigma^0,\omega)}<\sigma_x>_{\alpha(\sigma'^0, \omega')}\, ,
\end{equation}
where $<\cdot>_{\alpha}$ denotes the (thermal) average with respect to
the pure state $\rho^{\alpha}$. The possible overlap values
can be only $\pm q_{EA}$ and both of these
outcomes must have positive probability of occurring (as 
$\sigma^0, \omega, \sigma'^0, \omega'$ vary independently to yield a
pair of replicas).  But the 
translation-ergodicity of each of the distributions from which ${\cal J}$,
$\sigma^0, \omega, \sigma'^0, \omega'$ are chosen implies
the same for the joint (product) distribution of 
$({\cal J},\sigma^0,\omega,\sigma'^0, \omega')$.  Thus, the fact
that the overlap $Q$ is a translation-invariant function 
implies that Q must be constant for almost all
realizations, leading to a contradiction.
This argument can be immediately extended to any countable number
of pure state pairs. $\, \diamond$

\medskip

This proof also shows that if LNE does not occur
(and the limiting pure states $\alpha(\sigma^0,\omega)$ 
have nonzero $q_{EA}$), then
almost every (as $\sigma^0,\omega,\sigma'^0, \omega'$ vary)
overlap of the pair of pure states, $\alpha(\sigma^0,\omega)$ and
$\alpha(\sigma'^0, \omega')$, is zero.  This leads to:

\medskip

{\it Corollary}.  If LNE does not occur and $q_{EA} \ne 0$, 
then there must be an {\it uncountable\/} number
of pure states, with almost every pair (in the above sense) having overlap
zero.  

\medskip

This shows that, as stated in the introduction, nonequilibrium dynamics
provides important information on the structure of equilibrium pure
states.  It also suggests a dynamical test of the two-state picture:  search
for chaotic time dependence in $\nu_{t^*,\tau(t^*)}$.  If LNE does not
occur, then the two-state picture has been ruled out:  there must be
an uncountable number of states, almost all of which have overlap zero
(consistent with the results of Ref.~\cite{NS97}).  If
LNE does occur then neither the two-state nor the many-state
pictures have been ruled out.

How might one go about observing LNE in a spin glass,
where, unlike the ferromagnet, one doesn't know
what a domain wall looks like?  Here one can use
the clustering property
that characterizes pure states in general.  
E.g., a truncated 2-point correlation function of the
form $<\sigma_x \sigma_0>-<\sigma_x><\sigma_0>$ 
approaches 0 as $|x|\to\infty$
if the averaging is done in a pure state, but not otherwise.
In the current context a pure state average corresponds
to a dynamical average using $\nu_{t^*,\tau}$ with
$\tau\ll t^*$.  In principle one could evaluate
this correlation function numerically for $t^*\gg\tau\gg|x|\gg 1$;
if it does not approach zero for increasing values
of these parameters, that would constitute a clear signal of the
occurrence of LNE.  

An important consequence of Theorem 2 is that LNE must also 
occur at positive temperature
in the $2D$ uniform Ising ferromagnet and the random Ising ferromagnet
for $d<4$. (However, the argument for LNE in 
random ferromagnets is not entirely rigorous
as there is no complete analysis of interface 
pure states there$^{\cite{For}}$.
There is though a rigorous proof that for the SOS approximation,
these states do exist for $d \ge 4$ and do not for $d < 4$;
see Refs.~\cite{BK,BK2}).  

Moreover, in the $2D$ homogeneous ferromagnet 
(on the square lattice) results on LNE can be extended
to $T=0$, in the sense there that $\sigma^t$ does not
converge as $t\to\infty$; in fact, {\it every\/} 
spin flips infinitely often. The proof 
(by contradiction) is
based on showing that if some spin (say at the origin) remained fixed
forever, then (by translation ergodicity and
spin flip symmetry) so would two spins of opposite sign
on the $x$ and $y$ axes.  But then 
there would be a domain wall passing through the rectangle determined by
these three spins and ``cutting off'' one of them. In every time
interval there would be a nonzero probability of the domain wall moving
to flip that spin, and so with probability one, it
eventually {\it would\/} flip.  For the full proof, 
see Ref.~\cite{NN}.  

It is also shown in Ref.~\cite{NN}
that for many systems $\sigma^t$ {\it does\/} 
converge to some limit at $T=0$. 
This is based on a very general argument that there can be only
finitely many flips at any site that strictly lower the energy.
Examples include spin 
glasses and random ferromagnets where the 
common distribution of the $J_{xy}$'s is continuous, and either
has a finite mean (as in most ordinary models) 
or else is sufficiently ``spread out'',
as in the highly disordered spin glass or ferromagnet$^{\cite{NS94}}$.
Other examples are homogeneous ferromagnets 
(or homogeneous antiferromagnets or $\pm J$ spin glasses) on lattices with an
odd number of nearest neighbors, such as the $2D$ hexagonal lattice
or a double-layered $2D$ square lattice.
(There are also systems, such as the $\pm J$ spin glass 
on the square lattice, where some spins flip only finitely
many times and some spins flip infinitely often$^{\cite{GNS}}$.)
In light of these results, we restrict the term LNE to $T>0$, 
since in the zero-temperature situations where
$\sigma^t$ converges, the limit configuration is 
typically only metastable rather than a ground state
and so equilibration has not really occurred.
In these systems one can define a dynamical order parameter,
related to the autocorrelation, that does {\it not\/} decay
to zero.

To further clarify the discussion of LNE for the ferromagnet,
we note that it is a phenomenon separate from the spontaneous formation 
(at positive temperature) of domains of, say,
the minus phase within the plus phase.  That is, 
on a timescale exponential in $L$, there
will form a domain containing the origin
(of size $L$) of the minus phase.
Similarly, for a finite system of size $L$, the entire system
will flip back and forth between the plus and minus phases
on a timescale exponential in $L$.  However, this
is different from LNE, which 
presumably takes place on time 
scales of some power of $L$.  Also, the spontaneous formation
of droplets described above cannot occur at $T=0$,
but as already discussed, in the $2D$ ferromagnet the
phenomenon of domain walls forever sweeping across 
any finite region persists at zero temperature.

Since the existence of LNE for all $T<T_c$ in the $2D$ Ising
ferromagnet may seem surprising, 
we present a possible physical mechanism for
this case which may also shed light on LNE in general. 
The initial spin configuration 
has (with probability one) no infinite domains.  As 
the configuration evolves, some domains shrink and others
coalesce.  So the origin should always be contained in a finite domain, whose
size will usually be
slowly decreasing, but sporadically will have
a large change either by coalescing or because a domain wall
passes through the origin and the identity of the domain changes.
Thus LNE is primarily the result of nonequilibrium domain wall motion
(driven by mean curvature) combined with the complex domain
structure resulting from the original quench. It is also 
consistent with phase separation (as would be expected
from equilibrium roughening arguments). In particular, the
occurrence of LNE does not preclude the divergence with $t$ of
the {\it mean\/} scale of the domain 
containing the origin (although for {\it fixed\/} 
$\sigma^0$ and $\omega$, there is a much more complex behavior,
as indicated above).

\subsection{LNE and chaotic time dependence}
\label{subsec:ctd}

We noted earlier in this section that LNE can occur 
in the context of either possibility (1) (the
fully averaged dynamical measure $\nu_{t^*}$ has a limit,
which is a mixed state) or (2) ($\nu_{t^*}$ does not converge).
That is, LNE {\it must\/} occur if possibility
(2) holds, but may or may not occur if possibility (1) holds.
We now explore further the distinctions between the two cases of LNE.

As described earlier, one way for possibility (1) to 
occur is if, for some choice
of $\tau(t^*)$, $\nu_{t^*,\tau(t^*)} \to \rho^{\alpha(\sigma^0,\omega)}$
for almost all $\sigma^0$ and $\omega$, where
$\rho^{\alpha(\sigma^0,\omega)}$ is some pure state.  But it could
also happen that for any choice of $\tau(t^*)<<t^*$, $\nu_{t^*,\tau(t^*)}$
never settles down to a single pure state --- so the system is usually
in a pure state locally, but the pure state forever changes.  
Nevertheless, a full average over the dynamics 
(i.e., letting $\tau = t^*$) still yields a single limit.  This is
to be contrasted with possibility (2), where even the fully averaged
measure never settles down.

To clarify these statements, we use the illustration of
the $2D$ homogeneous ferromagnet below $T_c$, where we know
LNE to occur by Theorem~2.  Suppose that possibility (1) occurs.
Then (for fixed $\sigma^0$ and large time) 
for approximately half of the dynamical realizations, 
a region of fixed lengthscale $L$ surrounding the origin
is in the up state (i.e., the pure Gibbs state $\rho^+$),
and for most of the other half the same region
is in the down state (the pure Gibbs state $\rho^-$), 
and this one-to-one ratio remains essentially fixed after
some timescale depending on $L$.  Then as $t^*\to\infty$,
$\nu_{t^*}\to\overline\rho$, where $\overline\rho$ is the 
mixed Gibbs state $(1/2)\rho^+ + (1/2)\rho^-$.
Nevertheless, in any given dynamical realization 
(with averaging done as usual after time $t^*-\tau$,
with $1<<\tau<<t^*$), the region never settles permanently
into either $\rho^+$ or $\rho^-$.

By contrast, if possibility (2) occurs, then even
the fully averaged dynamical measure $\nu_{t^*}$
forever changes.  This could happen (again for
fixed $\sigma^0$) if the random dynamics fails
to sufficiently ``mix'' the states (in which case one
has, given $\sigma^0$, some amount of predictive
power for determining from $\sigma^0$ the likely 
state of the system in the region for arbitrarily
large times $t^*$). This is conceivable because even though
$\sigma^0$ is globally unbiased between the plus and minus 
states, it does have fluctuations in favor of one or the
other state of order $\sqrt{(L^*)^2}$ on 
lengthscale $L^*$; with $L^*$ taken as an appropriate power 
of $t^*$, these fluctuations could (partially) predict the
sign of the phase at the origin at time $t^*$.
In possibility (1) on the other hand, there is
a greater
capability of the random dynamics to ``mix'' the states
which eventually destroys the predictive power contained in the 
fluctuations of the initial state.  

So there are really two kinds of non-equilibration, corresponding 
either to LNE in the framework of
possibility (1) (``weak LNE'') or else to LNE resulting from the
stronger possibility (2).
Because $\nu_{t^*}$ evolves deterministically according to an
appropriate master equation, its lack of a limit in possibility (2)
corresponds to the usual notion of deterministic chaos and can
thus legitimately be called chaotic time dependence (CTD)$^{\cite{FIN}}$.  
If weak LNE occurs, this term is not appropriate
because here the effect is due to the random dynamics.

In our discussion of LNE in the previous section, we do not
yet know which of the cases correspond to CTD
and which to weak LNE.  This remains a problem for future
investigation.  However, we note that the occurrence of 
LNE (but not CTD) in homogeneous ferromagnets
(on ${\bf Z}^d$) is implicit in the (nonrigorous for $d \neq 1$) analysis
of persistence exponents$^{\cite{BDG,DBG,DG,MBCS}}$
in the sense that the fraction of sites that remain in the
same phase from time $t_1$ to time $t_2$ tends to zero for
$1 << t_1 << t_2$.
On the other hand, the fact mentioned above that the $T=0$ analogue of
this phenomenon is {\it lattice dependent\/} appears to have 
gone unnoticed.

\section{Time-averaged dynamical measures}
\label{sec:fixed}

We return to one issue 
that needs discussion, and will
be treated in greater detail elsewhere.
If LNE occurs (e.g., because only a single pair of pure states is present), 
would a (long) time average of, say, the spin at the origin (or,
for the ferromagnet, the magnetization in a finite region), give zero?
The answer is:  not necessarily.  It could be that, after
long times, the system has spent roughly equal amounts of time
in both states, in which case the usual time average$^{\cite{BY}}$
(or a discrete average over equally spaced times) {\it would\/}
approach zero.  But it could also happen that, after any long time,
the system has spent significantly more of its life in one or the other
state (which itself would change with the observational timescale).
In other words (still using the example of a two-state system),
at any long time the weights of the two states, as defined
by a dynamical measure involving a 
fixed ${\omega}$ and an average over uniformly
spaced times, could be different from $1/2$, and moreover
they may change with time.  This is analogous to an equilibrium 
phenomenon discovered by K\"ulske$^{\cite{Kuelske}}$ and
seems to be exactly what occurs in homogeneous ferromagnets, as reported in
recent numerical studies$^{\cite{DG}}$.  
To get a zero average in this situation one would 
need to average over a {\it sparse\/} sequence of
increasingly separated times.  

\section{Summary}
\label{sec:summary} 

We have presented a rigorous approach
to the dynamics of infinite spin systems which introduced
various dynamical measures on the spin configurations,
and considered whether they evolve 
into pure states.  We showed that in the
case of the EA Ising spin glass with broken spin-flip symmetry, 
one (or both) of two interesting things must
happen:  either LNE occurs, where the system never
settles into any pure state (i.e., domain walls forever pass
through any finite region, causing it forever to change its
pure state)\footnote{LNE may seem somewhat analogous to ``weak ergodicity
breaking'' (see, e.g., Ref.~\cite{Bouchaud})
in which the system forever moves to different 
metastable states with increasing
lifetimes.  There are important differences, however:  that scenario 
explicitly requires the system to remain within a single pure state 
for all time (and unlike LNE is supposed to be able to occur even 
if there is only one pure state), and further makes 
no reference to what is occurring in 
real space.}, or else there exist uncountably many
pure states, with almost every pair having zero overlap.

We proved that the union of the 
basins of attraction of all pure states (again,
if broken symmetry occurs) forms a set of measure
zero in configuration space; i.e., almost every starting configuration
is on a boundary between (several or possibly all) pure states.
While this is also true for the ferromagnet, it obviously is still easy
to prepare that system in a pure state.  
But this result has serious dynamical
consequences for the spin glass, and not only for 
deep quenches.  Because of the possibility of chaotic temperature
dependence \cite{BM,FH88}, it is potentially 
relevant even for small temperature
changes made slowly.  Experimentally observed slow relaxation
and long equilibration times in spin glasses may therefore be
a consequence of small (relative to the system) domain size
and slow (possibly due to pinning) motion of domain walls.

More generally, we have argued against 
a common viewpoint that pure state multiplicity is
irrelevant to the dynamics of infinite (or very large)
systems on finite timescales.
A system need not --- and in several cases, does not --- 
spend all of its time in a single pure state, 
even locally.  Because of this, it is also not necessarily 
true that ``absolutely broken ergodicity'' --- i.e., the 
presence of more than one pure state separated by infinite
barriers --- implies that time averages and Boltzmann
averages must disagree (or equivalently, that the limits $N\to\infty$
and $t\to\infty$ cannot commute).  Both averages can be zero
if for each state the (infinite) system locally spends
(roughly) equal amounts of time in it and its global flip,
as discussed in Section~\ref{sec:fixed}.
The other possibility is that after almost any long time,
the system has spent more of its life in one or the other
state (which itself would change with the observational timescale).
If this is the case, the averages should disagree, but due to a mechanism 
different from the standard one.  Further development of these 
ideas, and a discussion of their application to experiment,
will be presented in a future paper.

\medskip

{\it Acknowledgments.\/}  We thank Anton Bovier, David Huse  and
Christof K\"ulske for useful
comments.  CMN thanks WIAS, Berlin for its hospitality and DLS thanks
the Aspen Center for Physics.

\renewcommand{\baselinestretch}{1.0}

\end{document}